\documentclass[12pt]{scrreprt}
\newcommand{\Author}{Michael Stewart}
\newcommand{\Title}{Algorithmic Diversity for Software Security}
\newcommand{\Subtitle}{}
\title{\Title}
\subtitle{\Subtitle}
\date{\today}
\author{\Author}

\usepackage[T1]{fontenc}
\usepackage[utf8]{inputenc}
\usepackage{graphicx, epstopdf, wrapfig, float}    
\usepackage{alltt, amsmath, cite, multirow, tabularx, ragged2e, booktabs, caption, mdwlist, dcolumn}
\usepackage{caption,algorithm,algpseudocode}
\usepackage{hyperref}                  
\hypersetup{
    pdfauthor={\Author},    
    pdftitle={\Title},      
    unicode=false,          
    pdftoolbar=true,        
    pdfmenubar=true,        
    pdffitwindow=false,     
    pdfstartview={FitH},    
    colorlinks=true,       
    linkcolor=black,          
    citecolor=black,        
    filecolor=magenta,      
    urlcolor=cyan,          
    }

\begin{document}
\maketitle

\begin{abstract}
\section*{Abstract}
Software diversity protects against a modern-day exploits such as code-reuse attacks. When an attacker designs a code-reuse attack on an example executable, it relies on replicating the target environment. With software diversity, the attacker cannot reliably replicate their target. This is a security benefit which can be applied to massive-scale software distribution. When applied to large-scale communities, an invested attacker may perform analysis of samples to improve the chances of a successful attack \cite{unibus}. 

We present a general NOP-insertion algorithm which can be expanded and customized for security, performance, or other costs. We demonstrate an improvement in security so that a code-reuse attack based on any one variant has minimal chances of success on another and analyse the costs of this method. Alternately, the variants may be customized to meet performance or memory overhead constraints. Deterministic diversification allows for the flexibility to balance these needs in a way that doesn't exist in a random online method.
\end{abstract}


\section{Motivation}
Software diversification for security increases the cost an attacker has to pay for a successful attack. When a code-reuse attack is crafted on a sample of diversified software, it no longer has a guarantee of success when applied to a diversified version of the same program \cite{unibus}. The attacker is partly uninformed about the low-level structure of future code snippets it wants to reuse. The defender is partly uninformed  about which variant will be analysed by the attacker and which will be targeted for attack. To minimize the expected chance of successful attack, vulnerable code should be diversified between builds. In the absence of specific knowledge of future vulnerabilities, a general strategy is to maximize diversity throughout the entire executable.

Producing a diverse community can be decomposed into two tasks: a method of producing diverse variants and a method of selecting the order in which variants are distributed for use. The first task is where all the diversity takes place. Previous techniques perform that task in an online manner during distribution. In these online systems, diversification is entangled with distribution. The result is random diversification with no shared memory between compilation sessions. These complications make it difficult to optimize diversity. A more powerful deterministic diversification process can be achieved by separating the diversification task from compile-time. Deterministic methods provide better diversity and efficiency. The deterministic algorithm uses any existing diversification technique, such as NOP-insertion or function shuffling, and creates a set of output variants which are randomly distributed later at compile-time.

We will explain how to evaluate diversity in the context of security. Then we will introduce a deterministic diversification algorithm and a framework for comparing it with alternate methods.

\section{Previous work}
\label{previous work}
Shacham introduced the return-oriented attack\cite{shacham07}, a code-reuse method in which code snippets called "gadgets" are used to perform arbitrary operations for the attacker. These attacks take advantage of the liberal geometry of x86 assembly language: heterogeneous instruction length, jumps to arbitrary offsets, and dense instruction encoding create many overlapping instruction sequences. If an arbitrary executable address is called, it will run an instruction which may not have been present in the original program design. Return-oriented attacks use chain together gadgets that end in a RETURN instruction by modifying the stack. Newer techniques use alternate methods to combine gadgets, such as a focus on gadgets ending in a JUMP instruction rather than a RETURN \cite{jop2011}.
These methods circumvent the write-or-execute W$\oplus$X defence by executing only code that exists on the target system. They are also capable of arbitrary operations when carefully crafted. Shacham showed that the common Unix \emph{libc} library contained Turing-complete sets of gadgets, and most libraries present in modern systems do as well.

Code-reuse attacks rely on analysis of a file to catalogue functional gadgets and their locations within the executable. Software diversity functions as a defence against these attacks by invalidating a gadget catalogue for other executables in the population. Gadgets the attacker relies on may be moved, removed, or replaced with alternate functions. Franz' diversifying compiler (multicompiler) automates this process at compile time \cite{unibus}.


\section{Evaluating Diversity}
\label{Evaluating Diversity}
A diversification process can be thought of as a set of possible output states (variants) as well as the frequency and order in which each variant is produced. An ideal method will produce a community of executables with these qualities: The number of variants should equal the size of the population. That is, each variant should be used once. Additionally, each variant should share as few common gadgets as possible with every other variant. Entropy estimates can show the relationship between these parameters and the probability of successful attack.

\subsection{Entropy}
\label{Entropy}
For a random process, an entropy estimate reflects the amount of information required to describe the output of the process \cite{shannon1948}. For security purposes, this is useful for estimating the amount of information that an attacker will have to collect in order to optimally predict the results of the process. For example, if after studying some small number of sample executables, an attacker observes a pattern that useful gadgets tend to appear at the front of the memory space, the attacker will know to ``aim'' for that area in the future. By increasing entropy, we ensure that the attacker is left without these types of clues or at least require the attacker to invest in gathering more samples in order to find one.

 $$ S = -\sum_c [ P(c) * log_b( P(c) ) ] $$

$P(c)$ is the probability of a particular "state" occurring. A valid definition is such that the greater the total entropy of our system, the less information an attacker is expected to learn by studying samples of the population.

Jackson defines a "state" as a particular output state of a program from the compiler \cite{jackson2012}. Under this model, $P(c)$ is the probability of any particular output state of the compiler. All of the options available to the compiler at compile time will determine this probability. In a stochastic compilation setting, this can become complex and make $P(c)$ difficult to identify.

For the purposes of this work, each the possible output states is called a variant, and a ``state'' is the case of a particular gadget existing in another variant. This definition reflects more of the attacker's perspective than the defender's. It results in a simpler analysis and preserves the purpose of the entropy measure. $P(c)$ is the probability that a "functionally equivalent  gadget", one that has the same effect on processor state \cite{jackson2012}, exists at the same location in another variant. It is as if an attacker has studied one and targeted the other. If we have special knowledge of an attacker's habits or the code in question, the model can be modified to account for them. Otherwise, we simplify the model by assuming a strict worst-case scenario:
\begin{itemize*}
    \item The attacker has only chosen one gadget for their attack.
    \item The attacker can choose any gadget present.
\end{itemize*}

For a population size $N$, $c$ is the state referring to a particular gadget. $P(c)$ equals the number of other binaries that contain that gadget from the same starting position (or an equivalent) divided by $N-1$. In real-world situations, any attack will use one or more gadgets, so this $P(c)$ is an upper bound on the true probability of a successful attack. 

In the absence of more specific information about what gadgets an attacker will use, we want $P(c)$ across all gadgets to be as uniform and small as possible. 
This maximizes entropy; then the least amount of information is available to attack analyses. Our diversification algorithms are the tools that determine $P(c)$. In terms of the parameters mentioned earlier, this is achieved when the diversification algorithm with variety tuned to the size of the target population and maximum diversity between variants.

Compile-time diversification, which frequently uses iterative passes and Bernoulli trials, creates many exponential processes. It cannot control which variant is produced at compilation time. So it will always be less secure than a deterministic diversification algorithm.

\subsection{Variants and Population Size}
\label{Variants and Population Size}
A population size parameter reflects how many builds are expected to be made.  If a developer creates 600 variants but uses only 300, they might have made 300 more diverse variants or 300 at the same diversity while saving some costs along the way.

On the other hand, if 300 variants have been used, but an additional 300 binaries need to be distributed, the developer has two choices: use each existing variant twice, or \emph{ad hoc} generate a complementary set of variants numbered 301 through 600. The first choice maintains the performance and variety of the smaller population size. The second choice effectively doubles the variety of the population and increases diversity, at a theoretical cost to efficiency. In both cases, the total population isn't distributed uniformly. If the correct population size was known ahead of time, the variants could have been distributed more efficiently.

\section{Pattern-Based Approach}
\label{Pattern-Based Approach}
The general method for creating transformation patterns is 
\begin{enumerate}
    \item Identify scope of possible outputs
    \item Restrict output to those that achieve desired variety, diversity, or cost
\end{enumerate}

Once this is done, the patterns can be applied at will. Where a random compile-time approach will repeat certain decisions a random number of times, a pattern-based approach can use each pattern precisely the number of times desired. 

\begin{figure}[h] \centering
\includegraphics[width = .4\textwidth]{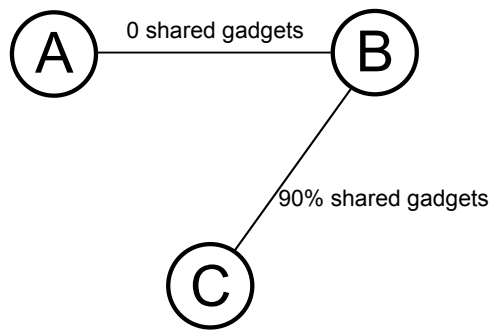} 
\caption{A and B alone are more diverse than A, B, and C}
\end{figure}

The strength of pre-computed patterns is the ability to selectively choose which output states will be used. Any possible output can be filtered out, used once, multiple times, or saved for future use. The following algorithms focus on selectively reducing the number of variants in order to maximize diversity.

\begin{figure}[H] \centering
\includegraphics[width = .6\textwidth]{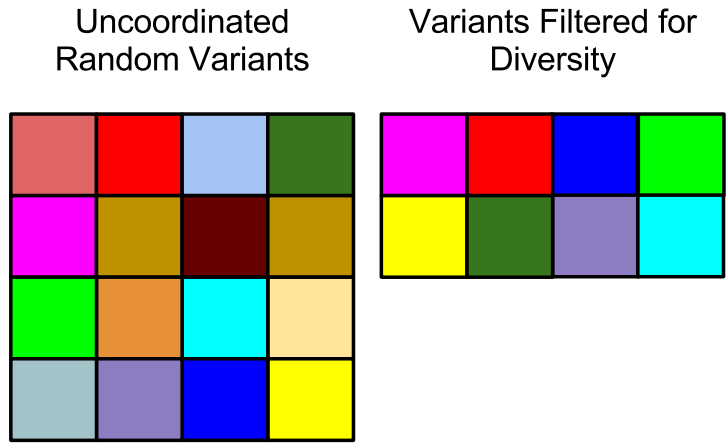} 
\caption{Reducing variety to increase diversity}
\end{figure}

\subsection{Function Permutation}
\label{Function Permutation}
Function permutation chooses a random permutation of the order in which functions are written to the resulting executable at link time. The advantages of permutation patterns are that it doesn't use additional disk space, and is reported to have negligible effect on performance \cite{jackson2012}. The disadvantage is its limited variety.

\begin{figure}[h] \centering
\includegraphics[width = \textwidth]{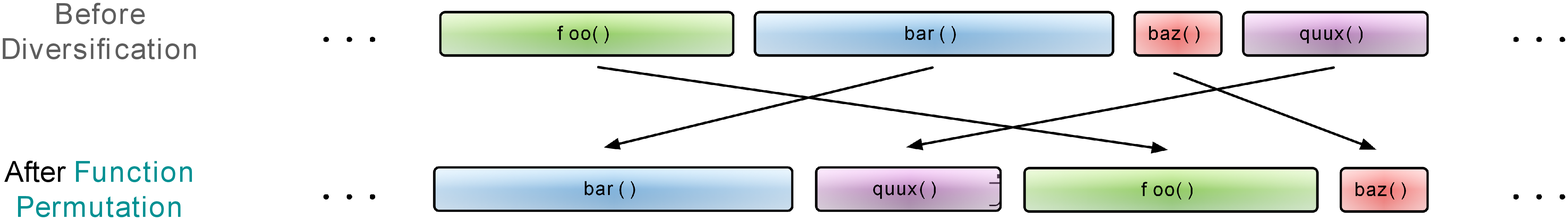} 
\cite{jackson2012}
\end{figure}

A compile-time function permutation might use the same permutation multiple times, or use less-diverse variants together e.g. "2 0 3 4 5 6 1" and "2 0 3 5 6 1 6". To maximize diversity, one or both of these variants would be excluded. While reducing the variety allowed, diversity is improved. Maximal diversity is created by rotation of functions. The result is $F$ number of patterns, where $F$ is the number of functions in the source program. While there are $F!$ possible permutations, only sets of $F$ permutations do not contain the same function in the same position multiple times.  In the best case scenario, all functions contain disjoint sets of gadgets. Algorithm \hyperref[Algorithm Permutation Patterns]{``Permutation Patterns''} produces a maximal diversity function permutation community.

    \begin{algorithm}
	    \label{Algorithm Permutation Patterns}
		\caption{Permutation Patterns}
		\begin{alltt}
		    Identify the number of functions F    
		    Create a single permutation of function indices
		    Create (F-1) additional permutations by rotating
		      the previous permutation
		    Randomly permute the order of this list of permutations
		\end{alltt}
	\end{algorithm}

    \begin{figure}
	\caption*{Permutation patterns example for a program with $7$ functions}
	\begin{verbatim}
	    Starting with single permutation of function indices:
	        2 0 3 4 5 6 1 
	    Make complimentary patterns via rotation: 
	    Pattern 0:    2 0 3 4 5 6 1 
	    Pattern 1:    0 3 4 5 6 1 2 
	    Pattern 2:    3 4 5 6 1 2 0 
	    Pattern 3:    4 5 6 1 2 0 3 
	    Pattern 4:    5 6 1 2 0 3 4 
	    Pattern 5:    6 1 2 0 3 4 5 
	    Pattern 6:    1 2 0 3 4 5 6 
	
	    Then randomly permute the order of those variants so they
	    can be popped uniformly at compile time:
	    Pattern 0:    4 5 6 1 2 0 3 
	    Pattern 1:    6 1 2 0 3 4 5 
	    Pattern 2:    2 0 3 4 5 6 1 
	    Pattern 3:    1 2 0 3 4 5 6 
	    Pattern 4:    0 3 4 5 6 1 2 
	    Pattern 5:    3 4 5 6 1 2 0 
	    Pattern 6:    5 6 1 2 0 3 4
	\end{verbatim}
	\end{figure}

\subsection{NOP Insertion}
\label{NOP Insertion}
Compile-time NOP insertion uses a rate parameter $p$ to a Bernoulli trial for each instruction of the source program. If the trial is successful, a NOP is inserted in front of that instruction \cite[p56]{jackson2012}. This method has an exponentially larger chance of affecting gadgets later in the file, but creates a lot of random "noise" in the middle of each. A NOP insertion creates a shift in the instructions that follow. It has a potential diversifying effect on every gadget that follows it.

Another defence technique is to use NOP insertions to break existing gadgets. NOPs can be inserted between intended instructions without affecting the program semantics. When gadgets begin in one intended instruction and continue into another, a NOP inserted between them can remove the gadget without breaking the way the program was intended to behave \cite{movingtarget}.

\begin{figure}[h] \centering
\includegraphics[width = .8\textwidth]{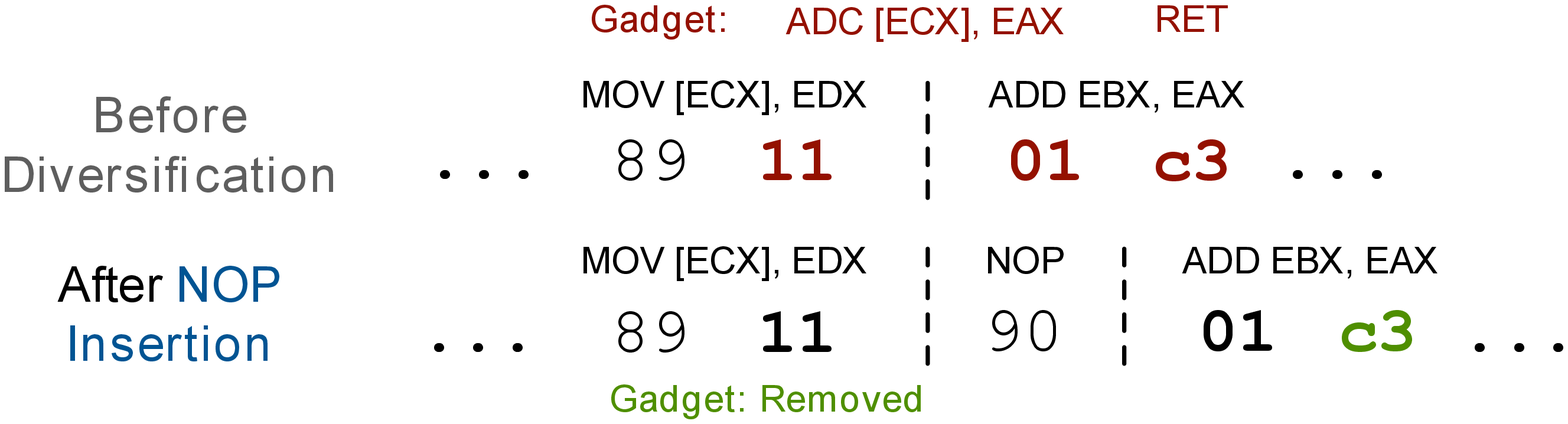} 

\cite{jackson2012}
\end{figure}

\subsubsection{NOP Padding}
If a NOP pad of sufficient length is added to the beginning of a file, each gadget from the original will be displaced into the gadget that follows it. This produces binaries with disjoint gadget sets. When produced iteratively from a minimal base variant, maximum diversity is achieved for a minimum number of NOPs inserted. Maximal diversity is guaranteed between all previous variants by the monotonic difference between each pair and the global constraints described below. 

\begin{algorithm}
    \caption*{NOP Padding}
    \label{Algorithm NOP Padding}
    \begin{alltt}
        Start with an unmodified build
        For the target population size P,
            Insert a single new pad in front of the previous build
            Emit this as the newest build
    \end{alltt}
\end{algorithm}

\begin{figure}
	\caption{\{A, B, C, D\} are gadgets, aligned by offset. N is a NOP pad. Note: There is no pad shown at the end of these files, but each should contain one for obfuscation.}
	\begin{verbatim}
	    File 1:    A B C D
	
	    File 2:    N A B C D
	
	    File 3:    N N A B C D
	
	    File 4:    N N N A B C D
	
	    ...
	\end{verbatim}
\end{figure}

NOP-padding works as a low-level equivalent to address space layout randomization or base address randomization \cite[p94]{movingtarget}, except that it can also be used with low-level operating system binaries. A few complications have to be accommodated to optimize diversity.
Real binaries have multiple gadgets that are functionally equivalent. We need to avoid shifting one gadget into the position previously held by an equivalent gadget. Doing so requires extensive gadget analysis that can identify all gadgets that exist with or without NOP insertions and their functions.

\begin{figure}
\caption{Noisy NOP insertion. Files 1 and 2 contain gadget C accessible from offset 3.}
	\begin{alltt}
	    Offset:    0 1 2{\bf 3} 4    
	    
	    File 1:    A B N{\bf C} D
	
	    File 2:    N A B{\bf N} C D
	    
	    File 3:    N N A B N C D
	\end{alltt}
\end{figure}

A NOP already exists at offset 2 in file 1. File 2 is generated with a naive NOP pad at the beginning. Because of the existing NOP in the middle of file 1, gadget $C$ is accessible in file 1 and 2 from offset $3$. An additional pad, shown in file 3, pushes gadget $B$ into this offset, diversifying acceptably. The resulting pad length (in this case, 2) is a minimum-size pad which will be added iteratively to each pattern.

\begin{figure}[H]
Function alignment is an extreme case of the noise problem. It reduces the effect of a single base NOP pad on anything outside the first function.
	\centering
	\begin{tabular}{l c | l }
	            & Function 1    & Function 2    \\
	    File 1: &   A B C N N N &{\bf D} ... \\
	    File 2: &   N A B C N N &{\bf D} ... \\
	    File 3: &   N N A B C N &{\bf D} ... \\
	    File 4: &   N N N A B C &{\bf D} ... \\
	\end{tabular}
	\caption*{Gadget D is in a second function, aligned by the compiler}
\end{figure}


\subsubsection{NOP Noise}
NOP instructions inserted randomly throughout the file provide their own advantages. They can break existing gadgets\cite{jackson2012} and make later gadgets more resilient to analysis. If an attacker somehow identifies the global offset provided by the base pad at the beginning of the file, they won't know the offset of \emph{every} gadget in the rest of the file.

\begin{figure}[H]
Noise can be added to the padding algorithm above. Each noisy NOP added to a random location creates complications for diversity as described above. The following rules avoid those complications: Noisy NOPs present in a base pattern should be preserved in subsequent patterns to prevent duplicate gadgets. They can be moved backward. They can't be moved forward or removed.

	\captionof{figure}{Duplicates occur at offset 3}
	\begin{alltt}
	    Offset:      0 1 2{\bf 3} 4    
	    
	    Base File:   A B N{\bf C} D
	
	    Removing the internal N leaves a duplicate
	    Alt 1:       N A B{\bf C} D
	    
	    Migrating it forward also does    
	    Alt 2:       N A B{\bf C} N D
	    
	    Migrating it toward offset zero does not    
	    Alt 3:       N A N{\bf B} C D
	\end{alltt}
\end{figure}

\subsubsection{Noise Cost}
\label{Noise Cost}

\begin{figure}[H]
Noisy gadgets have the side-effect of decreasing the variety of resulting transformations. Consider all variants possible with a bound of two NOPs, on a simple binary with two gadgets A and B:

	\centering
	\begin{tabular}{l | l}
	    Pad-Only      & Pad+Noise    \\
	    \hline
	    A B           & A B \\
	    N A B         & N A N B  \\
	    N N A B       & \\
	\end{tabular}
	\caption{Noisy insertions reduce variety of transformations}
\end{figure}

With the noisy NOP inserted in the second set, for any given number of future NOP insertions, the noisy set will have at least one fewer variant in it. For a high enough noise rate, gadgets that occur later in the executable will begin to cluster toward the end of the file. Entropy estimates for these communities will be lower than for smaller noise rates.

\label{Algorithm NOP Patterns} 
\begin{algorithm}
    \centering
	\caption{NOP Patterns}
	\begin{alltt}
	  For population size P and noise rate N:
	     Starting with an unmodified build:
	          Identify the minimum base pad
	          Identify all sets of equivalent gadgets
	          Create black-list of offsets for each gadget which
	            has downstream equivalents
	      Initialize an output list of patterns with the unmodified build        
	      For (P-1) iterations:
	          Copy the last pattern in the output list
	          Increment the base pad by the minimum amount
	          Add noise NOPs into the pattern according to N
	          If a gadget G from the unmodified build is now 
	            at a blacklisted offset:
	              Add an additional noise NOP before G
	          Add the new pattern to the output list
	      Randomly permute the output list
	\end{alltt}
\end{algorithm}



NOP padding requires a linear number of NOP insertions per build, while making and preserving noise decreases the amount of diversity provided at the cost of memory space. The disadvantage of these NOP methods is that they require more disk space for the resulting file depending on the population size. The advantage is that they support large populations, limited by available memory.

\section{Methods}
\label{Methods}
To implement and test these diversification algorithms, we extended an existing diversifying compiler to implement the \hyperref[Algorithm Permutation Patterns]{``Permutation Patterns''} and \hyperref[Algorithm NOP Padding]{``NOP Padding''} algorithms. Note that in the ``NOP Padding'' algorithm, NOP-instructions are stored between output variants and the number of NOPs per variant increases monotonically.

Minimal analysis was performed on the source program, and then only for the function shuffling method which requires the number of existing functions in the program. Then a sequence of permutation or NOP output variants was created. The order in which all variants were to be used was decided by a random shuffle on the order in which they were stored, and the storage then treated as a queue. When it was time for the compiler to perform a build, a pattern was pulled from the queue and applied during link time. We used LLVM 2.9 and Clang together with the Gold linker to apply program-wide transformations at link time.

We experimentally identified a minimum-byte pad of 60 on a toy Hello-World program. 
A base pattern is created and stored to a list. Then additional patterns are iteratively constructed from the last pattern in the list. 

For consistency with previous multicompiler tests \cite[p87]{jackson2012}, we created diversified communities of SQLite3.6.9. We used the same minimum-byte pad of 60 rather than automate analysis and identify an optimal pad size. We implemented noisy insertions in NOP patterns in a very simple way: NOPs are inserted with a rate parameter as in the compile-time method, but previous NOPs are preserved and not moved between patterns.

\section{Results}
\label{Results}


\subsection{Diversity Comparison}
\label{Diversity Comparison}

For each transformation type, We created a set of 25 diversified versions of SQLite3.6.9. The noisy pattern-based NOP-insertion used a noise rate of 5\%. We also created a "vanilla" version using Clang's default settings.
We used Jackson's \emph{Survivor} analysis tool to identify surviving gadgets between each combination of binaries in the diversified group \cite[p128]{jackson2012}. This uses a conservative algorithm which over-estimates the number of surviving gadgets between executables.

The output can be formatted as a raw count of how many pairs of files contain a common gadget or as an aggregate number of unique gadget-location pairs that exist among any number of files. The raw count reflects the simple chance of an uninformed attacker finding success with a random gadget in a single attack. The aggregate count shows a rough estimate of the spread across binaries per common gadget.

For example, the population generated with noisy NOP patterns contained 311 instances where a gadget in one file survived in a second file. The aggregate count of surviving gadgets was 301. In this case, 5 gadgets survived in 3 different files, while 296 survived across only 2.

Contrast this with the population generated through compile-time NOP insertion at a 50\% rate: 621 common instances occurred, but 600 aggregate instances were made. Not only did more common gadgets survive, but those that did so remained more spread among the population. NOP patterns retained about 50\% as many gadgets as this non-pattern approach.

\bigskip
\begin{minipage}{\linewidth}
\centering
\captionof{table}{Surviving Gadgets Counts} \label{Counts Table} 
\begin{tabularx}{\linewidth}{c c c}\toprule[1.5pt]
\label{securitytable}
\bf Method & \bf Number of Gadgets Found & \bf Aggregate Gadgets \\\midrule
Pattern-based NOP-Insertion & \multirow{2}{*}{311} & \multirow{2}{*}{301} \\
Pad + Noise & & \\
\hline
Pattern-based NOP-Insertion & \multirow{2}{*}{388} & \multirow{2}{*}{382} \\
Pad Only & & \\
\hline
Compile-time NOP05 & 3706 & 2871 \\
\hline
Compile-time NOP25 & 1465 & 1345 \\
\hline
Compile-time NOP50 & 621 & 600 \\
\hline
Compile-time NOP75 & 732 & 718 \\
\hline
Compile-time NOP100 & 1580 & 1416 \\
\bottomrule[1.25pt]
\end{tabularx} 
\par 
\bigskip
\end{minipage}

In the padded NOP pattern community, 379 of the 382 unique gadgets were shared between only 2 binaries. The remaining three were the same gadget, \texttt{'add esp, 0x1C; ret'} in three different offsets. For each of these three offsets, the gadget was present across three binaries.
In the padded noisy pattern community, 296 of the 301 unique gadgets were shared between only 2 binaries. There were five instances of 3-variant gadgets.
The compile-time communities had larger sets of shared gadgets: NOP-05 had 215 3-variant gadgets, NOP-25 had 61 3-variant gadgets. 

\subsubsection{Evaluation}
In both cases, the deterministic approach can achieve the same or better diversity using the same or fewer NOP insertions than the random methods. The general counts in Table \ref{securitytable} show this intuitively. An entropy estimate allows for direct comparison between methods. Where $N$ is the community size, and each gadget is spread across $b$ builds, the entropy is $S$: 
\begin{align*}
 S &= -\sum_c [ P(c) * log{ P(c) } ] \\
 S &= -\sum_c [ \frac{b_c}{N} * log{\frac{b_c}{N}} ] 
\end{align*}

When $b=1$, the gadget in question only exists at that location in one build. In this case, $P(c)$ represents the success probability of an attacker using a gadget location that only exists in a single build (the one the attacker analysed.) We expect this to be zero, because we assume the attacker doesn't get to target the same build they analysed. Unfortunately, $log(0)$ is undefined and we cannot assume which build the attacker will target. So we assume equal probability that the attacker will target each existing build \emph{including} the ones analysed: $P(c) = \frac{b_c}{N}$. By assuming this ignorant stance about the attacker's true target, the entropy score reflects information presented to an equally ignorant attacker. They are a general representation and valid for attackers unaware of which build they are targeting.

The entropy estimate also attempts to weigh the effect of having a few gadgets common to many builds with having many gadgets present in fewer builds. The entropy calculation must reflect a comparison between many low-spread gadgets and few high-spread gadgets. The calculations presented here use a simple assumption and are not meant to address the trade-off between low-spread and high-spread gadgets. They are meant to address rough differences between sets of both. The magnitudes are not as important as how scores compare to each other. 

For example, the noisy NOP pattern community has higher total entropy than the noiseless NOP pattern community in spite of more frequent 3-build gadgets. Even if some other weight was given to say "attackers prefer 3-build gadgets over 2-build gadgets", the deterministic communities would still have far fewer remaining vulnerabilities than the randomly-diversified communities. In summary, deterministic patterns are more secure. 


\begin{table}
\centering
\captionof{table}{Surviving Gadgets Spread} \label{Spread Table} 
\begin{tabularx}{\linewidth}{c c c}\toprule[1.5pt]
\label{gadget spread table}
\multirow{2}{*}{\bf Method} & \multirow{2}{*}{\bf Builds per Gadget Found} & \bf Shannon Entropy \\
                            &                                              & \bf (bits) \\
\midrule
Pattern-based NOP-Insertion & 2-build gadgets: 296 & \multirow{2}{*}{15022} \\ 
Pad + Noise & 3-build gadgets: 5 & \\

\midrule
Pattern-based NOP-Insertion & 2-build gadgets: 379 & \multirow{2}{*}{15019} \\ 
Pad Only & 3-build gadgets: 3 & \\

\midrule
\multirow{4}{*}{Compile-time NOP05} & 2-build gadgets: 2592 & \multirow{4}{*}{14895} \\ 
 & 3-build gadgets: 215 & \\
 & ... & \\
 & 9-build gadgets: 1 & \\

\midrule
\multirow{4}{*}{Compile-time NOP25} & 2-build gadgets: 1275 & \multirow{4}{*}{14976} \\ 
 & 3-build gadgets: 61 & \\
 & 4-build gadgets: 8 & \\
 & 6-build gadgets: 1 & \\

\midrule
\multirow{3}{*}{Compile-time NOP50} & 2-build gadgets: 586 & \multirow{3}{*}{15010} \\ 
 & 3-build gadgets: 13 & \\
 & 4-build gadgets: 1 & \\

\midrule
\multirow{3}{*}{Compile-time NOP75} & 2-build gadgets: 710 & \multirow{3}{*}{15005} \\ 
 & 3-build gadgets: 6 & \\
 & 4-build gadgets: 2 & \\

\midrule
\multirow{4}{*}{Compile-time NOP100} & 2-build gadgets: 1343 & \multirow{4}{*}{14972} \\ 
 & 3-build gadgets: 63 & \\
 & ... & \\
 & 7-build gadgets: 1 & \\

\bottomrule[1.25pt]
\end{tabularx} 
\end{table}

\clearpage
\subsection{Costs}
\label{Costs}
\subsubsection{Performance}
For performance tests, we used the same binaries used in the diversity comparison, in addition to 25 more, for a total 50 diversified versions of SQLite3.6.9 for each transformation type. We used a comparison benchmark Tcl script provided by the SQLite developer \cite{sqlitespeedtest}, modified it to test only SQLite, then ran this script on each variant.

\bigskip
\begin{minipage}{\linewidth}
\centering
\captionof{table}{Performance Overhead} \label{Performance Table} 
\begin{tabularx}{\linewidth}{c p{4cm} p{5cm}}\toprule[1.5pt]
\bf Method & \bf \centering {\% Synchronization (variance)} & \bf \% No Synchronization (variance) \\\midrule
No transformations & 0 (0) & 0 (0) \\
\hline
Pattern-based NOP-Insertion & \multirow{2}{*}{6.758 (4.81)} & \multirow{2}{*}{8.256 (3.90)} \\
Pad + Noise & &\\
\hline
Pattern-based NOP-Insertion & \multirow{2}{*}{ -1.544 (1.35) } & \multirow{2}{*}{ -0.383 (0.894) } \\
Pad Only & &\\
\hline
Compile-time NOP05 & 2.654 (1.05) & 3.119 (1.1) \\
\hline
Compile-time NOP25 & 1.926 (1.38) & 3.376 (1.29) \\
\hline
Compile-time NOP50 & 3.402 (1.14) & 4.586 (.999) \\
\hline
Compile-time NOP75 & 6.123 (2.91) & 6.771 (1.89) \\
\hline
Compile-time NOP100 & 4.463 (1.47) & 6.069 (1.91) \\

\bottomrule[1.25pt]
\end{tabularx} \par
\bigskip
	Synchronization is an internal parameter of SQLite.
\end{minipage}

\subsubsection{File Size}

For 50 binaries of each compile-time transformation, we took the mean file size. For each of the pattern-based transformations, the file size generated is a function of the population size (in this case, 50). By contrast, compile-time diversification space costs \emph{are not} a function of population size.
Link-time optimization with no NOPs inserted created 615588-byte files. Clang's default optimizations add about 8Kb of space to the vanilla file, for 623780 bytes.

\bigskip
\begin{minipage}{\linewidth}
\centering
\captionof{table}{File Size Overhead} \label{File Size Table} 
\begin{tabularx}{\linewidth}{c c c}
\toprule[1.5pt]
\bf Method & \bf File Size (bytes) & \bf \% Overhead \\\midrule
No transformations & 623780 & 0\\
\hline
Pattern-based NOP-Insertion & \multirow{2}{*}{694340} & \multirow{2}{*}{+11.31} \\
Pad + Noise &\\
\hline
Pattern-based NOP-Insertion & \multirow{2}{*}{619684} & \multirow{2}{*}{-.0657} \\
Pad Only & &\\
\hline
Compile-time NOP05 & 625146 & +0.210 \\
\hline
Compile-time NOP25 & 667930 & +7.078 \\
\hline
Compile-time NOP50 & 719058 & +15.27 \\
\hline
Compile-time NOP75 & 771316 & +23.65 \\
\hline
Compile-time NOP100 & 824484 & +32.18 \\

\bottomrule[1.25pt]
\end{tabularx}
\end{minipage}

\section{Conclusions}
\label{Conclusions}

NOP patterns displayed more diversity per byte applied during compilation. The 50\% community had 600 gadgets and an entropy score of 15010, about 4\% performance cost and 15\% additional disk space. In contrast, padding-only NOP patterns had 385 gadgets, entropy score 15019, 2\% performance \emph{improvement}, and slightly less disk space than the undiversified build. The community of NOP patterns with noise showed much higher performance overhead, 11\%, but had fewer gadgets remaining than either of the previous two.

This implementation used regular x86 NOPs, which had a negative impact on performance. Alternate NOP instructions should mitigate most of this performance overhead \cite{jackson2012}. Thus, for the community of programs generated by the tested compile-time method, a more diverse community can be created for almost no cost by using a pattern-based method.

Resulting diversity or variety can be reduced to support larger populations or smaller cost requirements. It is tunable in the same ways compile-time methods are. For these tests, maximal diversity was attempted without analysis of the program code. The minimum-pad size and noise rate used in noisy NOP patterns were unoptimized. A smaller pad or noise rate might have achieved the same effect at a smaller cost. NOP-padding complications that would remove the remaining gadgets were ignored. 


\section{Future Work}
\label{Future Work}
Testing on NOP patterns with very high noise rates could illustrate \hyperref[Noise Cost]{theoretical diversity costs} of noisy internal NOP insertions. Gadgets may cluster toward the end of files in these communities.

NOP patterns might benefit from a combination with other transformations. For a given cost budget and diversity target, NOP patterns support a certain population size, $N$. If permutation patterns on its own supports a population of 2 for the same parameters, then the combination of NOP patterns and permutation patterns could maintain the original requirements for a population of size $2*N$.

\chapter*{Acknowledgements}
Thanks to Prof. Michael Franz and the Secure Systems Laboratory for their help and support.
Andrei Homescu and Todd Jackson provided invaluable insight to C++, the Linux build process, and the SSL Multicompiler.
Dr. Per Larson provided guidance, inspiration and illustrations for this project.
Thanks to Dr. Stefan Brunthaler for advice and support in Linux and python, and Mason Chang for encouragement and style.

\label{Bibliography}
\bibliography{systems}
\bibliographystyle{ieeetr}

\end{document}